\documentclass[]{article}
\usepackage{lmodern}
\usepackage{amssymb,amsmath}
\usepackage{ifxetex,ifluatex}
\usepackage{fixltx2e} % provides \textsubscript
\ifnum 0\ifxetex 1\fi\ifluatex 1\fi=0 % if pdftex
  \usepackage[T1]{fontenc}
  \usepackage[utf8]{inputenc}
\else % if luatex or xelatex
  \ifxetex
    \usepackage{mathspec}
  \else
    \usepackage{fontspec}
  \fi
  \defaultfontfeatures{Ligatures=TeX,Scale=MatchLowercase}
\fi
\IfFileExists{upquote.sty}{\usepackage{upquote}}{}
\IfFileExists{microtype.sty}{%
\usepackage{microtype}
\UseMicrotypeSet[protrusion]{basicmath} % disable protrusion for tt fonts
}{}
\usepackage[margin=1in]{geometry}
\usepackage{hyperref}
\hypersetup{unicode=true,
            pdftitle={Ontology-based multi-agent system to support business users and management},
            pdfauthor={Dejan Lavbič, Olegas Vasilecas and Rok Rupnik},
            pdfborder={0 0 0},
            breaklinks=true}
\usepackage{natbib}
\usepackage{longtable,booktabs}
\usepackage{graphicx,grffile}
\makeatletter
\def\maxwidth{\ifdim\Gin@nat@width>\linewidth\linewidth\else\Gin@nat@width\fi}
\def\maxheight{\ifdim\Gin@nat@height>\textheight\textheight\else\Gin@nat@height\fi}
\makeatother
\setkeys{Gin}{width=\maxwidth,height=\maxheight,keepaspectratio}
\IfFileExists{parskip.sty}{%
\usepackage{parskip}
}{% else
\setlength{\parindent}{0pt}
\setlength{\parskip}{6pt plus 2pt minus 1pt}
}
\providecommand{\tightlist}{%
  \setlength{\itemsep}{0pt}\setlength{\parskip}{0pt}}
\ifx\paragraph\undefined\else
\let\oldparagraph\paragraph
\renewcommand{\paragraph}[1]{\oldparagraph{#1}\mbox{}}
\fi
\ifx\subparagraph\undefined\else
\let\oldsubparagraph\subparagraph
\renewcommand{\subparagraph}[1]{\oldsubparagraph{#1}\mbox{}}
\fi

\let\rmarkdownfootnote\footnote%
\def\footnote{\protect\rmarkdownfootnote}

\usepackage{titling}

  \title{Ontology-based multi-agent system to support business users and
management}
    \pretitle{\vspace{\droptitle}\centering\huge}
  \posttitle{\par}
    \author{Dejan Lavbič, Olegas Vasilecas and Rok Rupnik}
    \preauthor{\centering\large\emph}
  \postauthor{\par}
    \date{}
    \predate{}\postdate{}

\usepackage{booktabs}
\usepackage{longtable}
\usepackage{array}
\usepackage{multirow}
\usepackage[table]{xcolor}
\usepackage{wrapfig}
\usepackage{float}
\usepackage{colortbl}
\usepackage{pdflscape}
\usepackage{tabu}
\usepackage{threeparttable}
\usepackage{threeparttablex}
\usepackage[normalem]{ulem}
\usepackage{makecell}

\usepackage{amsthm}

\theoremstyle{definition}

\theoremstyle{definition}

\theoremstyle{definition}

\theoremstyle{remark}

\let\BeginKnitrBlock\begin \let\EndKnitrBlock\end
\begin{document}
\maketitle

\begin{quote}
\textbf{Dejan Lavbič}, Olegas Vasilecas and Rok Rupnik. 2010.
\href{https://doi.org/10.3846/tede.2010.21}{\textbf{Ontology-based
multi-agent system to support business users and management}},
\href{https://www.tandfonline.com/toc/tted21/current}{Technological and
Economic Development of Economy \textbf{(TEDE)}}, 16(2), pp.~327 - 347.
\end{quote}

\section*{Abstract}\label{abstract}
\addcontentsline{toc}{section}{Abstract}

For some decision processes a significant added value is achieved when
enterprises' internal Data Warehouse (DW) can be integrated and combined
with external data gained from web sites of competitors and other
relevant Web sources. In this paper we discuss the agent-based
integration approach using ontologies (DSS-MAS). In this approach data
from internal DW and external sources are scanned by coordinated group
of agents, while semantically integrated and relevant data is reported
to business users according to business rules. After data from internal
DW, Web sources and business rules are acquired, agents using these data
and rules can infer new knowledge and therefore facilitate decision
making process. Knowledge represented in enterprises' ontologies is
acquired from business users without extensive technical knowledge using
user friendly user interface based on constraints and predefined
templates. The approach presented in the paper was verified using the
case study from the domain of mobile communications with the emphasis on
supply and demand of mobile phones.

\section*{Keywords}\label{keywords}
\addcontentsline{toc}{section}{Keywords}

Decision support, agent, multi-agent system, ontology, data warehouse,
information retrieval, business rules, business process management

\section{Introduction}\label{introduction}

There is a growing recognition in the business community about the
importance of knowledge as a critical resource for enterprises. The
purpose of knowledge management is to help enterprises create, derive,
share and use knowledge more effectively to achieve better decisions, to
increase of competitiveness and to decrease the number of errors. In
order to run business effectively an enterprise needs more and more
information about competitors, partners, customers, and also employees
as well as information about market conditions, future trends,
government policies and much more. There are several products and
technologies available on the market that support advanced Business
Process Management (BPM) \citep{trkman_process_2007} and advanced
decision support. Enterprises expect these applications to support wide
range of functionalities - analysing customer profiles, building and
analysing business strategies, developing customer-specific products,
carrying out targeted marketing and predicting sales trends. Amount of
documents in the Web, enterprise data repositories, and public document
management systems with documents are rapidly growing. This huge amount
of data is managed in some extent, but knowledge workers, managers, and
executives still have to spend much of their working time reading dozens
of various types of electronic documents spread over several sources in
process of making decisions. There is just too much information to
digest in a daily life. The tremendous amount of documents that is still
growing has far exceeded the human ability for comprehension without
intelligent tools. Different applications within information systems
(IS) that support wide range of functionalities need to be integrated in
order to provide the appropriate level of information support. One of
the prominent approaches for IS integration is the use of ontologies and
Multi-Agent Systems
\citep{fuentes_heterogeneous_2006, soo_cooperative_2006, dzemydiene_multi-layered_2009}.

The approach presented in this paper is targeted towards using
ontologies for several tasks, where emphasis is on using business rules
(BR) approach for interoperability between business user and IS. By
introduction of BR approach business users do not have to be fully
familiar with the technology to manipulate the common understanding of a
problem domain in a form of ontology and therefore enabling agents to
execute defined analyses models. The use of ontologies in Multi-Agent
System (MAS) environment enables agents to share a common set of
concepts about contexts, user profiles, products and other domain
elements while interacting with each other. Agents can exploit the
existing reasoning mechanisms to infer derived contexts from known
contexts, to make decisions and to adapt to the environment, current
status, and personal setting of the user. The purpose of this paper is
to present the approach of integration of several information resources
for Decision Support in Enterprises using agent-oriented approach based
on ontologies. The goal of our research is to minimize the gap between
business users and agents as special type of application systems that
perform tasks in their behalf. The intention was to apply BR approach
for ontology manipulation in MAS. Ontology used in our Multi-Agent
System for Decision Support in Enterprises (DSS-MAS) was divided into
task and domain ontologies while business users were enabled to
manipulate them directly in a user friendly environment without
requirement of detailed technical knowledge.

The remainder of this paper is structured as follows. First we present
some background in the following section \ref{background} with emphasis
on agents, ontologies and related work with clear definition of the
problem and solution proposal. Next, in section \ref{dss-mas}, we
introduce our case study of integrated Multi-Agent environment from the
domain of mobile communications with emphasis on architecture and the
roles of agents and ontologies. The case study is focused in one of the
mobile operators and furthermore oriented to supply and demand of mobile
phones. After presentation of system architecture and decomposition of
ontology of every agent from DSS-MAS will be presented in detail.
Details of case study implementation will be given in section
\ref{case-study}. Finally the last section \ref{conclusion} presents
conclusions.

\section{Some background on decision support, multi-agent systems and
ontologies}\label{background}

\textbf{Decision support systems (DSS)} have evolved significantly and
there have been many influences from technological and organizational
developments \citep{shim_past_2002}. DSS once utilized more limited
database, modelling, and user interface functionality, but technological
innovations enabled more powerful DSS functionality. DSS once supported
individual decision makers, but later DSS technologies are applied to
workgroups or teams, especially virtual teams. The advent of the Web has
enabled inter-organizational DSS and has given rise to numerous new
applications of existing technology as well as many new decision support
technologies themselves. Internet facilitates access to data,
information and knowledge sources, but at the same time, it threatens to
cognitively overload the decision makers. Authors in
\citep{vahidov_decision_2004} claim that internet technologies require a
new type of decision support that provides tighter integration and
higher degree of direct interaction with the problem domain. Based on
that they propose a generic architecture for dynamic and highly complex
electronic environments where DSS's should be situated in the problem
domain. \citet{chen_integrated_2001} conducted an interesting research
about integrated interactive environment for knowledge discovery from
heterogeneous data resources. Their work is grounded on acquiring,
collecting, and extracting relevant information from multiple data
sources, and then forming meaningful knowledge patterns. The proposed
system employs common DW and OLAP\footnote{OnLine Analytical Processing
  (OLAP)} techniques to form integrated data repository and generate
database queries over large data collections from various distinct data
resources.

\textbf{Multi-Agent Systems (MAS)} offer a new dimension for cooperation
and coordination in an enterprise. The MAS paradigm provides a suitable
architecture for a design and implementation of integrated IS,
especially DSS. With agent-based technology a support for complex IS
development is introduced by natural decomposition, abstraction and
flexibility of management for organisational structure changes
\citep{kishore_enterprise_2006}. The MAS consists of a collection of
autonomous agents that have their own goals and actions and can interact
and collaborate through communication means. In a MAS environment,
agents work collectively to solve specific enterprises' problems. MAS
provide an effective platform for coordination and cooperation among
multiple functional units in an enterprise. The research on agents and
MAS has been on the rise over the last two decades. The stream of
research on IS and enterprise integration
\citep{lei_iips:_2002, kang_agent-based_2003, tewari_personalized_2003}
makes the MAS paradigm appropriate platform for integrative decision
support within IS. Similarities between the agent in the MAS paradigm
and the human actor in business organisations in terms of their
characteristics and coordination lead us to a conceptualisation where
agents in MAS are used to represent actors in human organizations.
Several approaches
\citep{tewari_personalized_2003, rivest_solap_2005, kishore_enterprise_2006, soo_cooperative_2006}
deal with agent support for integration and decision support. Research
in \citep{kishore_enterprise_2006} has shown that MAS paradigm provides
an excellent approach for modelling and implementing integrated business
IS. Authors within that research proposed a conceptual framework for MAS
based integrative business IS. Some promising results were also found in
\citep{soo_cooperative_2006}, where authors propose a cooperative MAS
platform to support the invention process based on the patent document
analysis. The platform allows the invention process to be carried out
through the cooperation and coordination among software agents delegated
by the various domain experts in the complex industrial R\&D
environment.

Today, semantic technologies based on \textbf{ontologies} and inference
are considered as a promising means towards the development of the
Semantic Web \citep{davies_semantic_2006}. In the field of Computer
Science and Information Technology (IT) in general ontology has become
popular as a paradigm for knowledge representation in Artificial
Intelligence (AI), by providing a methodology for easier development of
interoperable and reusable knowledge bases (KB). The most popular
definition, from an AI perspective, is given in
\citep{gruber_translation_1993} as follows: ``An ontology is an explicit
specification of a conceptualization'', where a conceptualization is
abstracted view of the world that we wish to represent for some purpose.
Ontologies can be considered as conceptual schemata, intended to
represent knowledge in the most formal and reusable way possible. Formal
ontologies are represented in logical formalisms, such as OWL, which
allow automatic inferencing over them. An important role of ontologies
is to serve as schemata or intelligent views over information resources.
Thus they can be used for indexing, querying, and reference purposes
over non-ontological datasets and systems, such as databases, document
and catalogue management systems. Because ontological languages have a
formal semantics, ontologies allow a wider interpretation of data that
is inference of facts which are not explicitly stated. In this way, they
can improve the interoperability of the conceptualization behind them,
their coverage of arbitrary datasets. Ontology can formally be defined
as specific sort of knowledge base and can be characterized as
comprising a 4-tuple \citep{davies_semantic_2006}

\begin{equation}
O = \left \langle C, R, I, A \right \rangle
\label{eq:O}
\end{equation}

Where \(C\) is set of classes representing \emph{concepts} we wish to
reason about in the given domain (\texttt{Offer}, \texttt{Finding},
\texttt{Phone}, \texttt{Customer}, etc.). \(R\) is set of
\emph{relations} holding between those classes
(\texttt{Message\ hasRecipient\ Actor}). I is a set of \emph{instances},
where each instance can be an instance of one or more classes and can be
linked to other instances by relations
(\texttt{Nokia\ is\ A\ PhoneBrand};
\texttt{Finding309\ hasValue\ 11,23}). \(A\) is a set of \emph{axioms}
(\texttt{If\ a\ new\ customer\ buys\ is\ A\ Nokia\ E72,\ promotional\ discount\ of\ 10\%\ should\ be\ offered}).
It is widely recommended that knowledge bases, containing concrete data
(instance data or ABox) are always encoded with respect to ontologies,
which encapsulate a general conceptual model of some domain knowledge,
thus allowing easier sharing and reuse of KBs. Typically, ontologies
designed to serve as schema for KBs do not contain instance definitions,
but there is no formal restriction in this direction. Drawing the
borderline between the ontology (i.e.~the conceptual and consensual part
of the knowledge) and the rest of the data, represented in the same
formal language, is not always trivial task. In our approach we include
instances as part of ontologies because instances we define are a matter
of conceptualization and consensus and are not only descriptions,
crafted for some purpose.

Related work on using ontologies in information systems for decision
support is extensive. Regarding the domain of DW and OLAP analyses
research has dealt with Document Warehousing \citep{tseng_concept_2006}
where extensive semantic information about the documents is available
but still not fully employed as in traditional DW. The use of ontologies
was found useful as a common interpretation basis for data and metadata.
Furthermore research has extended to Web DW
\citep{marotta_managing_2002} with the emphasis on managing the volatile
and dynamic nature of Web sources. Utilization of ontologies is also
addressed in Information Retrieval (IR) where it has been used for fuzzy
tagging of data from the Web
\citep{buche_fuzzy_2006, macias_providing_2007}, query construction tool
in semi-automatic ontology mapping \citep{suomela_user_2006} and
semantic based retrieval of information from the World Wide Web
\citep{shan_programmable_2003, garces_concept-matching_2006}. Use of
ontologies in Data Mining (DM) has also been considered in
\citep{bernaras_building_1996, zhou_rod_2002, singh_context-based_2003, cao_ontology_2004}
where ontology was used for representation of context awareness and
handling semantics inconsistencies. Ontologies have been widely used for
data, application and information integration in the context of domain
knowledge representation \citep{qiu_towards_2006}.
\citet{jovanovic_achieving_2005} concludes that the need for knowledge
sharing and interoperable KBs exists and the key element for achieving
interoperability are domain ontologies. In that approach XSLT\footnote{XSLT
  (XSL Transformations) is a declarative, XML-based language used for
  the transformation of XML documents into other XML documents}
transformation is used to enable knowledge interoperability. Authors in
\citep{vasilecas_ontology-based_2006} use ontologies for ontology based
IS development and they address the problem of automation of information
processing rules. There are also other approaches
\citep{fuentes_heterogeneous_2006, orgun_hl7_2006, clark_ontology_2007}
that use ontologies as knowledge representation mechanisms. Authors in
\citep{clark_ontology_2007} use a formal ontology as a constraining
framework for the belief store of a rational agent. The static beliefs
of the agent are the axioms of the ontology, while dynamic beliefs are
the descriptions of the individuals that are instances of the ontology
classes. Another work presented in \citep{fuentes_heterogeneous_2006}
also uses heterogeneous domain ontology for location based IS in a MAS
framework with the emphasis on context-aware MAS. They propose a global
ontology to let agents work with heterogeneous domains using a wireless
network and the intention is to provide customization about different
environment services based on user location and profile.

\subsection{Problem and proposal for
solution}\label{problem-and-proposal-for-solution}

The review of related work presented in this section pointed out that
modern DSS's changed quite substantially especially with the advent of
the Web and availability of extensive information in online
repositories. For managing complexity and integration issues with
decision support many approaches relied on MAS paradigm and used
ontologies as knowledge representation mechanism. The existing
approaches mainly focused on either supporting existing business
processes or improving decision support at some level of detail or
integration of several structured resources to achieve better decision
support. To our knowledge none of the approaches addressed the problem
of enriching data from internal data sources with unstructured data
found on internet. The interactivity of reviewed solutions is also
limited; meaning that business users are usually limited to small set of
parameters they can define to alter default behaviour of the system.
These user requirements are usually entered directly into the system and
no abstraction layers are provided as in business rules management
systems (BRMS) to enable users without technical skills to manipulate
the content.

This paper introduces a novel approach in integration of unstructured
information found in the Web with information available in several
internal data sources (e.g.~database, DW, ERP\footnote{Enterprise
  Resource Planning (ERP)}, etc.). The MAS paradigm with agents was used
for implementation purposes, mainly because related work pointed out
that it is a very appropriate solution for integration of business IS.
One of the reasons to choose agents is also modelling notion where
business users and agents are modelled in a very similar manner. Problem
of interaction between human actors and computer programs is also
addressed by introduction of ontologies as knowledge representation
mechanism. The approach presented in this paper is targeted towards
using ontologies for several tasks, where emphasis is on using BR
approach to ensure interoperability between business user and IS.
Ontology is used not only for every agent to represent the
interpretation of a problem domain but also for communication between
agents and business users. The use of ontologies in MAS enables agents
to share a common set of facts used in user profiles, product
descriptions and other domain elements, while interacting with each
other. With exploiting reasoning mechanisms new findings can be derived
from initially known facts and improve the KB by extending it with new
knowledge. To simplify this communication template system based on BR
was introduced to enable manipulation of knowledge within the system by
users with less technical skills and to control behaviour of individual
agent. The approach will be further explained in the following section.
The case study presented in this paper is from the domain of mobile
telecommunications. It is presented in detail (with the impact it has on
improving decision support within enterprise) in section
\ref{case-study}. In the domain of mobile communications that was used
for case study we had to define several tasks in DSS-MAS, needed for
decision support - OLAP analyses, DM, IR, context and profile
definition, notification, etc.

\section{DSS-MAS}\label{dss-mas}

\subsection{DSS-MAS architecture}\label{dss-mas-architecture}

\begin{figure}

{\centering \includegraphics[width=0.5\linewidth]{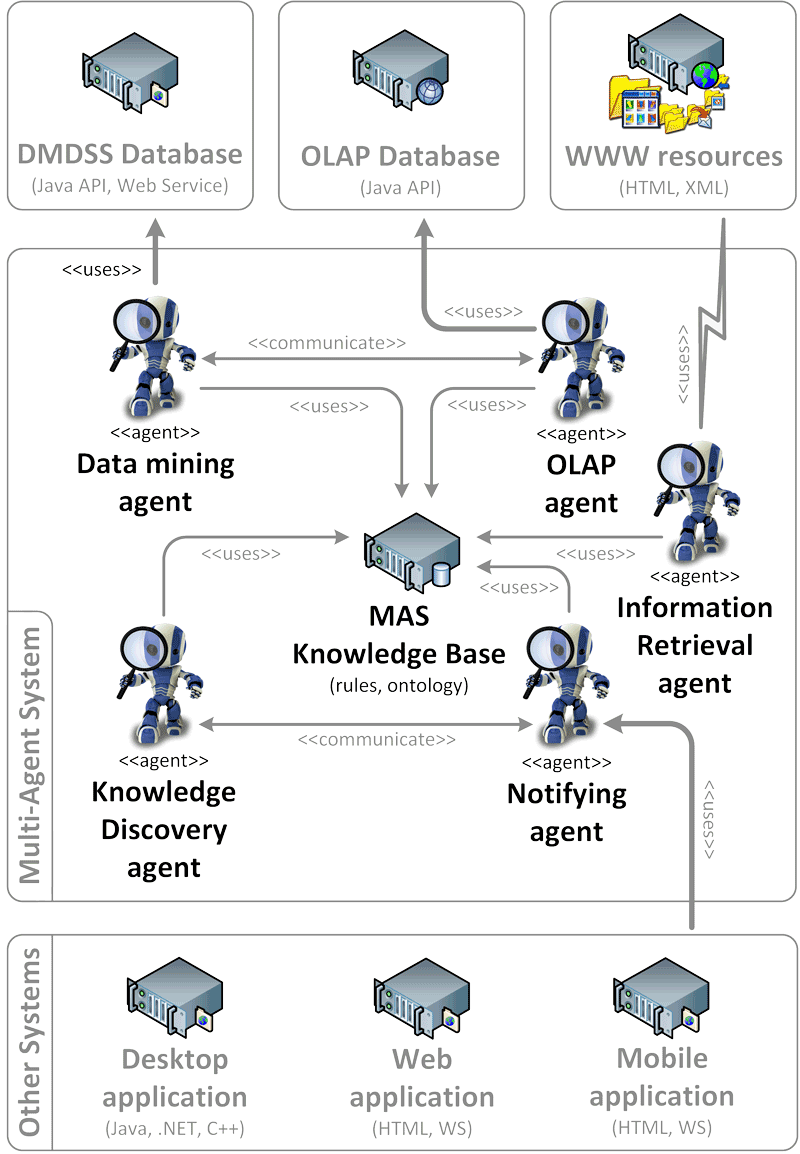}

}

\caption{Architecture of MAS used for Decision Support in Enterprises}\label{fig:architecture}
\end{figure}

DSS-MAS that we propose in this paper is introduced in Figure
\ref{fig:architecture}. The case study presented in this paper is from
the domain of mobile telecommunications and is based on business
environment and information resources from one of the mobile operators.
DSS-MAS is situated in the environment of several existing systems, like
Data Mining Decision Support System (DMDSS), or DW and various resources
available outside of an enterprise on the World Wide Web. Global goal
that agents in DSS-MAS should achieve is to support decision making
process while using existing systems for business analysis and employing
information from environment where enterprise resides. To support this
goal DSS-MAS includes several agent roles that are as following:
\textbf{Data Mining Agent (DMA)}, \textbf{OLAP Agent (OLAPA)},
\textbf{Information Retrieval Agent (IRA)}, \textbf{Knowledge Discovery
Agent (KDA)}, \textbf{Notifying Agent (NA)} and \textbf{Mobile Agent
(MA)}. The agents in DSS-MAS have both reactive and proactive
characteristics. Reactive are mainly due to responding to the
environment according to the model defined in the KB they use. Proactive
are due to their ability to learn from the environment and change the
initially defined KB to for example improve performance. Ontologies are
used as a main interconnection object for domain knowledge
representation, agent-to-agent communication and most important for
agent-to-business user communication. An important element of an
environment is the World Wide Web, where agents acquire information for
the purpose of decision making. Retrieved information is saved in a KB
and available for further employment for DM and DW analyses. All
information gathered from internal and external sources is considered by
KDA, where inference over several task ontologies used by individual
agents (DMA, OLAPA, IRA, etc.) is performed. Moreover the sub goal of
DSS-MAS is delivering of the right information at the right time and to
the right users. The system needs to be context aware and to consider
the relevant features of the business, i.e.~context information such as
time, location, and user preferences \citep{liao_constructing_2005}.
Business users in DSS-MAS are able to employ agents to perform tasks on
their behalf. For example, managers in enterprises have to request
reports from their systems - OLAP or from transactional databases, and
managers have to review reports every appointed period of time (day,
week, month, etc.). This task of information acquisition is predecessor
for decision making and is more or less straightforward - business user
sends a request for analyses and reviews the content according to some
Key Performance Indicators (KPI). KPI is simply a measure of performance
and is commonly used in enterprises to evaluate how successful they are.
In DSS-MAS tasks like this are automated and user participation is
reduced as much as possible. An initial analysis model (e.g.~OLAP or DM)
has to be captured in the ontology by business users, while execution
and optimisation is left for agents. Business users first define initial
parameters for analyses to be performed, while agents perform these
analyses and recommend improvements. When some action is required from
business user, he is notified and has the ability to act or change rules
of agent's execution.

To enable these functionalities we introduce ontologies as a mediation
mechanism for knowledge exchange between actors (agents and business
users) that cooperate in DSS-MAS. The following section will present the
structure and organization of ontologies we have used for the case
study.

\subsection{The role of ontology}\label{the-role-of-ontology}

\begin{figure}

{\centering \includegraphics[width=1\linewidth]{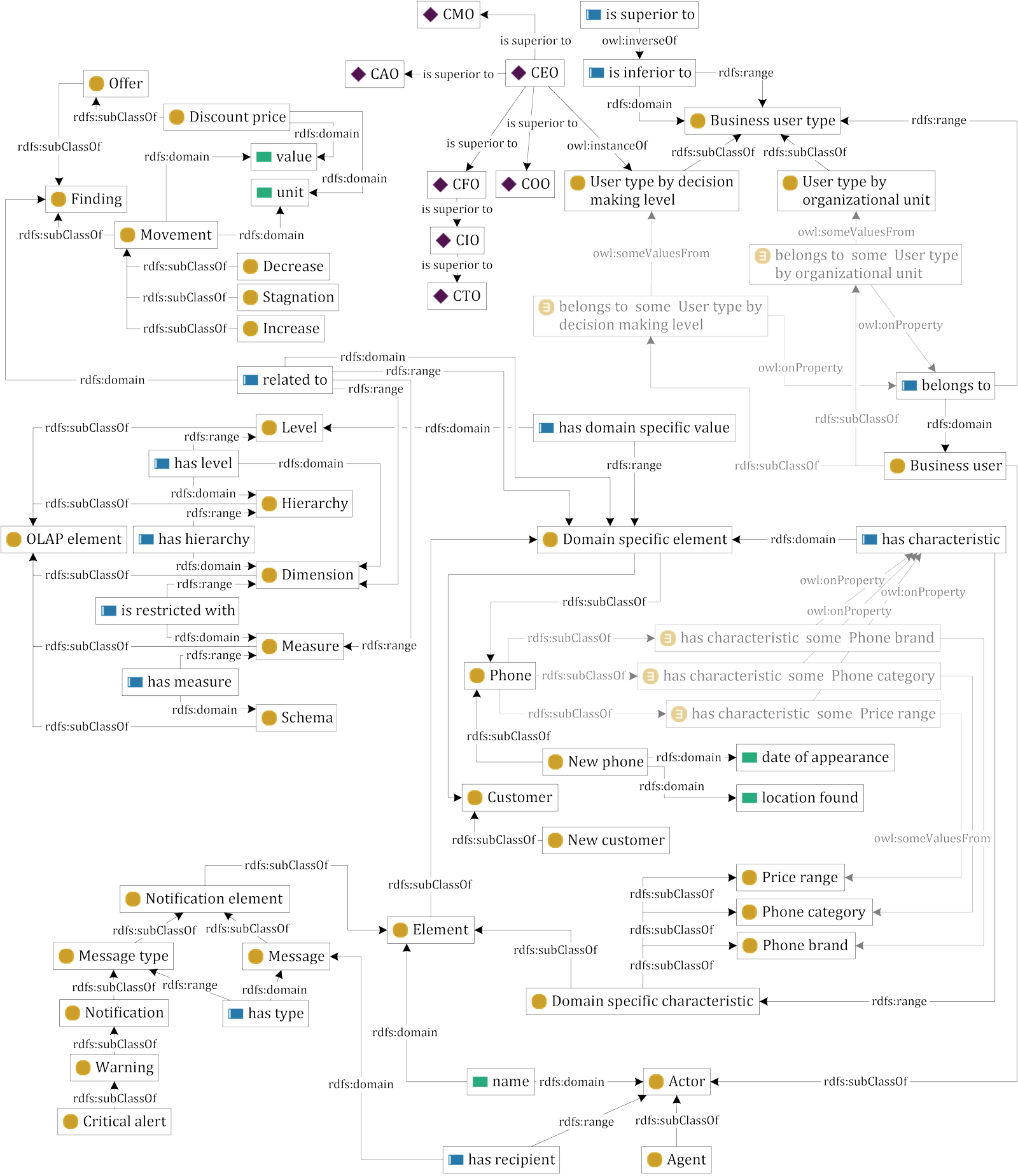}

}

\caption{Excerpt from intersection of several ontologies used in our case study}\label{fig:ontology-overview}
\end{figure}

According to \citet{guarino_formal_1998}, ontology can be structured
into different sub-ontologies - upper ontology, domain ontology, task
ontology and the application ontology. Following similar guidelines we
have defined upper ontology named \textbf{Common ontology} and combined
domain and task ontologies in \textbf{Notifying ontology},
\textbf{Information retrieval ontology}, \textbf{Data Mining and
Warehousing ontology} (see Figure \ref{fig:ontology-overview}). The
proposed clustering of ontologies is based on the common understanding
of the problem domain being defined in Common ontology. Every agent has
its own interpretation of a KB, which is a specialization of a Common
ontology with detail definition of knowledge required by individual
agent. Common ontology is limited to abstract concepts and it covers
reusable dimensions, which are primarily used by KDA. Task ontologies
specify concepts of notification, IR, DM and DW. Mobile communications
in our case is the domain of all task ontologies and the emphasis is on
supply and demand of mobile phones. As already mentioned, we have used
the knowledge management approach in our research where every agent has
knowledge about its own problem domain. In this case whenever new facts
about the common knowledge are discovered, which might be of interest
for other agents, they are updated to the common ontology.

The role of ontology in our approach is therefore twofold:

\begin{itemize}
\tightlist
\item
  knowledge representation mechanism used by agents and
\item
  common understanding of problem domain used for communication between
  business users and agents by utilizing business rules manipulation
  with introduced templates (see section \ref{mediation}).
\end{itemize}

Figure \ref{fig:ontology-overview} shows an excerpt from intersection of
several ontologies used in our case study. This part of ontology clearly
defines the common elements being used for communication between agents
and business users (domain specific elements such as phones, new phones
and customers, all described with domain specific characteristics). A
part of OLAP elements needed for conducting OLAP analyses is also
presented. Ontology also presents notification with taxonomy of various
warning levels and business users classification by organizational unit
and decision making level.

\subsection{The role of agents}\label{the-role-of-agents}

Our case study uses domain of mobile telecommunications as a platform
where we focus on the sales of mobile phones and their accessories.
Manipulation with internal data storage is handled by two types of
agents - OLAPA and DMA. They both have distinct tasks but still share
common goal -- periodically or on demand autonomously execute analyses
models. Business users at first define these models and describe them
with all required parameters (e.g.~search for anomalies in sales of
Nokia phones in last month period). The information about the execution
is stored in the ontology (based on business user preferences) or is
requested by another agent of the system. Business user preferences in
this context define the execution parameters about the analysis, for
example the period at which the analysis is performed (e.g.~perform
analysis every other day at 13:00). OLAPA has on firsthand
straightforward task of performing OLAP analyses on behalf of an agent
or a business user and reporting its findings back to the requesting
entity and all other entities that should be informed, according to the
business policy. Nevertheless OLAPA does much more - after each
execution it prepares the report for business user based on findings --
movements and KPIs. If certain finding is substantially different from
finding obtained in previous case further analysis is performed to
discover the reason of change by drilling down (more detailed) or up
(less detailed) the hierarchies and levels.

\begin{figure}

{\centering \includegraphics[width=0.8\linewidth]{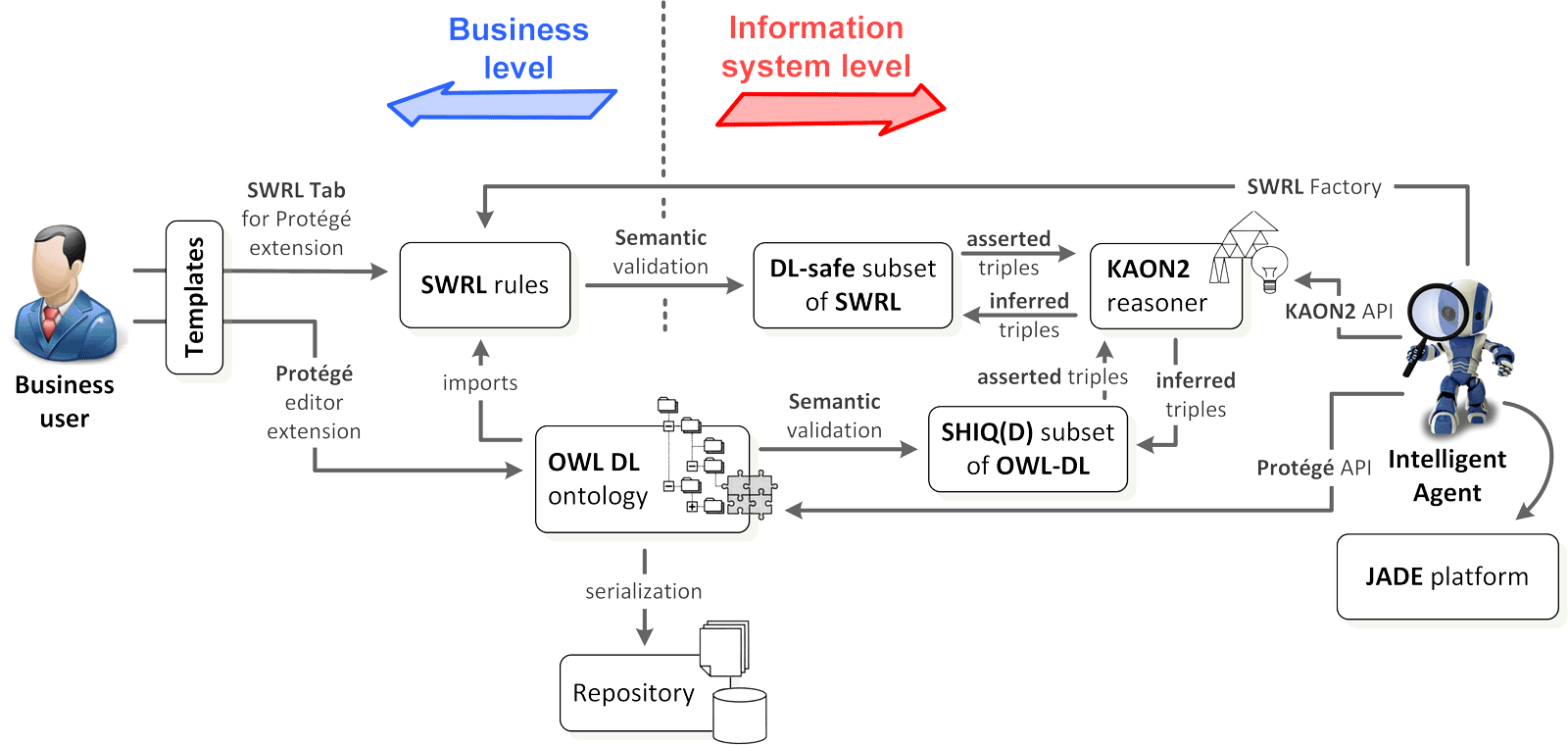}

}

\caption{Prototype of selected case study}\label{fig:case-study-prototype}
\end{figure}

The knowledge is acquired in ontology. Business users can change the
behaviour of agents by changing the ontology using graphical user
interface. This interface incorporates all logical restrictions defined
in ontology and does not allow users to enter unacceptable values and
the most important is that it does not require technical knowledge from
users. Previous our experiences have shown that business users have
great difficulties especially with setting the parameters required to
run DM and DW analyses models. So, user interface really has to be
friendly and intuitive. In our approach this issue was solved by
introducing the architecture depicted in Figure
\ref{fig:case-study-prototype} and using templates as further discussed
in section \ref{case-study}.

Nowadays Web-based information retrieval systems are widely distributed
and deeply analysed from different points of view. The main objective of
all such systems is to help users to retrieve information they really
need (obviously as quickly as it is possible)
\citep{garces_concept-matching_2006}. While the techniques regarding DW,
multi-dimensional models, OLAP, or even ad-hoc reports have served
enterprises well, they do not completely address the full scope of
existing problems. It is believed that, for the business intelligence
(BI) of an enterprise, only about \(20\%\) of information can be
extracted from formatted data stored in relational databases
\citep{tseng_concept_2006}. The remaining \(80\%\) of information is
hidden in unstructured or semi-structured documents. This is because the
most prevalent medium for expressing information and knowledge is text.
For instance, market survey reports, project status reports, meeting
records, customer complaints, e-mails, patent application sheets, and
advertisements of competitors now are all recorded in documents. For
that reason in DSS-MAS we introduced IRA for retrieval of data mainly
from the World Wide Web. The tasks that IRA performs in presented case
study can be grouped into three categories:

\begin{itemize}
\tightlist
\item
  identification of new online shops,
\item
  analysis of mobile phones presented online and
\item
  extending Data Warehouse with information found online.
\end{itemize}

First two tasks are concerned about the supply of mobile phones at
various online shops worldwide. Identification of new online shops is
conducted with web crawling and the use of several existing services on
the Internet, such as Google, Google Product Search and Bing. Not only
these internet resources are managed through ontology, but also rules
for text extraction are defined as rules which make all domain knowledge
available in IR ontology and not encoded in agent itself. More details
about implementation of DSS-MAS case study can be found in section
\ref{case-study}. Furthermore every shop found online is analysed to
identify unique patterns for searching phones. Search patterns include
guidelines for agents performing search at various web pages. They are
based on XQuery\footnote{XQuery is a query language that is designed to
  query collections of structured and semi-structured data} and regular
expressions. To search for phones at Google Products Search, the
following URL search pattern
\texttt{http://www.google.com/products?q=((\textbackslash{}w+\textbackslash{}s*)+)}
is used, accompanied with additional information for web scrapping of
required information (e.g.~price, availability etc.). Using these search
patterns IRA is searching through online shops and determines phones
with their market prices and stores this information into IR ontology to
be available for further knowledge inference by KDA. Information of
found phones is used to determine new market trends, enable price
comparison between competitors, facilitate possible inclusion in
enterprise's sales program, etc. One of the tasks that IRA also performs
is extending DW with information found online. While business users
perform OLAP analyses, they deal with only internal information about
the business, but in process of decision making other resources also
have to be examined, e.g.~news about the suppliers and competitors,
opinions about certain products and organisations, etc. IRA therefore
scans the DW dimension data (through hierarchies and levels) from DW
dimensional schema and uses this information for searching several
internet resources (news archives, forums, stock changes, Google trends,
etc.). When users review OLAP reports these data from the Internet is
also displayed according to their restrictions in dimensions. For
example, when business users are making decision whether to increase
support to Nokia or Sony Ericsson phones it only has reports about sales
of selected brand names from their market program. Using our approach
the user is provided with additional data that is found online and what
will make decision better founded. By this integration of internal and
external information users have integrated data source available that
they can query from single location.

KDA is an important element of DSS-MAS since it consolidates all
findings from IR, DM and DW and furthermore delivers derived findings to
NA. To employ inference capabilities over several ontologies the
enterprises' BR are essential. While business concepts are captured in
ontology, these concepts further have to be restricted to define
specific meaning. Generally BR are prepared by business users and also
some parts of BR in enterprises tend to change frequently; therefore we
introduced architecture for BR management (see Figure
\ref{fig:case-study-prototype} and further discussion in section
\ref{case-study}). Findings of KDA are presented as instances of
\texttt{Domain-specific-element} and \texttt{Findings} classes (see
ontology in Figure \ref{fig:ontology-overview}).

As it can be seen from Figure \ref{fig:architecture} NA represents an
interface to DSS-MAS for all external applications and business users.
The main role of NA is the information dissemination by simply
delivering the right information at the right time to the right users.
While in vast majority of today's applications users have to request the
information using so called ``pull model'' in our approach we
implemented the ``push model'', where information is proactively
delivered by agents to the user without a specific request. This is
achieved by making system context aware and considering the relevant
features of the business, i.e.~context information such as time,
location, position in the organisational hierarchy, etc.

All knowledge about notification is defined in Notifying ontology, where
every user has his own context defined and the position within
organisation across two dimensions - organisational unit
(e.g.~Marketing, Sales, Human resources, etc.) and decision making level
(e.g.~Chief Executive Officer (CEO), Chief Information Officer (CIO),
Chief Financial Officer (CFO), Chief Marketing Officer (CMO), Chief
Analytics Officer (CAO), etc.). According to that position rules for
delivery of several message types are defined. These message types range
from Notification to Warning and Critical alert. Each message also
addresses the domain of specific organisational unit, e.g.~when a new
mobile phone is found online at competitor's website, CMO and CAO have
to be notified. Organisational structure, as part of Notifying ontology,
also defines that both CMO and CAO are inferior to CEO therefore he is
also notified, but only in a case of a Critical alert. According to the
business user profile, notification can be sent using several
technologies from Windows Alert, e-mail, Really Simple Syndication
(RSS), Short Message Service (SMS), etc. These notification types are
also ordered by priority for each business user and according to this
type the content is also adapted.

Mobile agent is an example of an application that can reside on a mobile
device (e.g.~Personal Digital Assistant (PDA), mobile phone, etc.) and
uses resources of DSS-MAS through NA. The typical use case includes
sending mobile agent across network to DSS-MAS, where all needed
information according to owner context is collected and then the mobile
agent is returned back to originating location on a mobile device and
presents the collected data to business user. When the process of
acquiring data is in progress, business user does not have to be
connected to the network, he can just wait offline until mobile agent is
ready to return with the findings.

In the following section details about the case study implementation
will be presented with technologies used, templates for business rules
acquisition and presentation of one specific scenario from case study.

\section{Case study implementation and discussion}\label{case-study}

\subsection{Technology}\label{technology}

The selected language for ontology presentation is OWL DL
\citep{russomanno_expressing_2004}, since it offers the highest level of
semantic expressiveness for selected case study and is one of the most
widely used and standardised ontology language nowadays that has
extensive support in different ontology manipulation tools. Besides OWL
logical restrictions, Semantic Web Rule Language (SWRL) rules were also
used due to its human readable syntax and support for business rules
oriented approach to knowledge management \citep{horrocks_owl_2005}.
SWRL rules are stored as OWL individuals and are described by OWL
classes contained in the SWRL ontology. The use of SWRL enables storing
schema, individuals and rules in a single component, which makes
management much easier. SWRL rule form in a combination with templates
that is introduced in the following subsection b is very suitable for
knowledge formalization by business users that do not have extensive
technical knowledge.

The ontology manipulation interface for business users is based on
Protégé Ontology Editor and Knowledge Acquisition System
\citep{stanford_medical_informatics_protege_2006} and SWRL Tab
\citep{stanford_medical_informatics_swrltab_2006} for Protégé. It
enables entering OWL individuals and SWRL rules where a step further is
made towards using templates for entering information (see Figure
\ref{fig:case-study-prototype}). At the execution level KAON2 inference
engine is used to enable inference capabilities. Due to limitation of
\(SHIQ(D)\) subset of OWL-DL and DL-safe subset of SWRL language, before
inference is conducted, semantic validation takes place to ensure that
all preconditions are met. We selected FIPA\footnote{Foundation for
  Intelligent Physical Agents (FIPA)} compliant MAS platform
JADE\footnote{Java Agent DEvelopment Framework (JADE)} in DSS-MAS
because it offers broad range of functionalities and is most widely used
platform. This is due to very good support and availability of agent
framework, where a lot of common agents' tasks are already implemented
(i.e.~agent communication at the syntax level, agent management,
migration of agents, etc.). For Mobile Agent implementation an add-on
JADE-LEAP\footnote{Java Agent Development Environment-Lightweight
  Extensible Agent Platform (JADE-LEAP)} was used to support the
mobility of agents.

\subsection{Mediation with BR templates}\label{mediation}

Using templates with ontology, business logic is excluded from the
actual software code whereas the majority of data for templates is
acquired from ontology axioms and natural language descriptions in
ontology, while other templates are prepared by users with technical
knowledge. The main goal of using mediation with BR templates is to
enable acquiring knowledge from actual knowledge holders i.e.~business
users and enable transformation of this high-level knowledge into
information system level, where this data together with concepts from
business vocabulary can be directly used for inference purposes and
bring added value without any further programming by technically
educated users.

When acquiring new knowledge into the system from business users, the
process always starts with focusing on concepts of business vocabulary
that are persisted in a form of ontology. Users can freely traverse
through this information space, select concepts and further manipulate
all related information within the selected context. Altering and adding
new information is all time limited to formal definition of concepts
that is defined in ontology. For easier manipulation business user is
aided with template and business vocabulary, so BR building process is
simplified as it will be presented in detail in the following section.

Example \ref{exm:br-template} presents a BR template that is used for
definition of aggregation of findings or domain specific elements. The
user interface that is available is directly linked to ontology, where
constraints on classes, properties and individuals are considered in
realtime. This approach allows to minimize the risk of entering wrong
constraints. The DSS-MAS system supports entering of new statements in
several forms from simple IF-THEN form to decision table or decision
tree.

\BeginKnitrBlock{example}[BR template for general finding definition]
\protect\hypertarget{exm:br-template}{}{\label{exm:br-template} \iffalse (BR
template for general finding definition) \fi{} }

\begin{align*}
&\textbf{IF} \\
&\quad \text{Condition} = \left \{ \exists x : x \in \textbf{Domain specific element} \cup \textbf{Finding} \right \} \wedge \left | \text{Condition} \right | \geq 1 \\
&\textbf{THEN} \\
&\quad \text{Result} = \left \{ \exists y : y \in \textbf{Finding} \right \} \wedge \left | \text{Result} \right | \geq 1
\end{align*}
\EndKnitrBlock{example}

The following Example \ref{exm:template-rule} represents a BR that
states:
\texttt{If\ there\ exist\ two\ consequent\ increases\ of\ sold\ phones\ of\ the\ same\ phone\ brand\ and\ a\ new\ phone\ of\ this\ phone\ brand\ was\ found\ online\ within\ last\ 2\ weeks,\ then\ offer\ a\ promotion\ discount\ of\ 10\%\ on\ this\ new\ phone\ to\ all\ new\ customers.}

\BeginKnitrBlock{example}[Example of a rule, developed by using template]
\protect\hypertarget{exm:template-rule}{}{\label{exm:template-rule}
\iffalse (Example of a rule, developed by using template) \fi{} }

\begin{align*}
&\textbf{IF} \\
&\quad \underline{\text{First finding}} \text{ is } \textbf{Increase (Finding)} \text{ which } \{ \\
&\qquad \text{is } \textbf{related to } \underline{\text{first amount sold}} \text{ which is } \textbf{Measure (OLAP element)} \text{ AND} \\
&\qquad \text{is } \textbf{related to } \underline{\text{first date}} \text{ which is } \textbf{Dimension (OLAP element)} \text{ AND } \\
&\qquad \text{is } \textbf{related to } \underline{\text{first phone}} \text{ which is } \textbf{Phone (Domain specific element)} \text{ which } \{ \\
  &\qquad \quad \text{has } \textbf{characteristic } \underline{\text{brand}} \text{ which is } \textbf{Phone brand (Domain specific characteristic)} \\
&\qquad \} \\
&\quad \} \text{ AND} \\
&\quad \underline{\text{Second finding}} \text{ is } \textbf{Increase (Finding)} \text{ which } \{ \\
&\qquad \text{is } \textbf{related to } \underline{\text{second amount sold}} \text{ which is } \textbf{Measure (OLAP element)} \text{ AND} \\
&\qquad \text{is } \textbf{related to } \underline{\text{second date}} \text{ which is } \textbf{Dimension (OLAP element)} \text{ which } \{ \\
&\qquad \quad \text{is } \textit{greater than } \underline{\text{first date}} \\
&\qquad \} \text{ AND} \\
&\qquad \text{is } \textbf{related to } \underline{\text{second phone}} \text{ which is } \textbf{Phone (Domain specific element)} \text{ which } \{ \\
&\qquad \quad \text{has } \textbf{characteristic } \underline{\text{brand}} \text{ which is } \textbf{Phone brand (Domain specific characteristic)} \\
&\qquad \} \\
&\quad \} \text{ AND} \\
&\quad \underline{\text{Found phone}} \text{ is } \textbf{New phone (Domain specific element)} \text{ which } \{ \\
&\qquad \text{has } \textbf{characteristic } \underline{\text{brand}} \text{ which is } \textbf{Phone brand (Domain specific element)} \text{ AND} \\
&\qquad \text{has } \textbf{date of appearance } \underline{\text{found date}} \text{ which is } \textbf{Dimension (OLAP element)} \text{ which } \{ \\
&\qquad \quad \text{is } \textit{greater than now - 14 days} \\
&\qquad \} \\
&\quad \} \text{ AND} \\
&\quad \underline{\text{New customer}} \text{ is } \textbf{New customer (Domain specific element)} \\
&\textbf{THEN} \\
&\quad \underline{\text{Promotion discount}} \text{ is } \textbf{Discount price (Finding)} \text{ which } \{ \\
&\quad \text{is } \textbf{related to } \underline{\text{new customer}} \text{ AND} \\
&\quad \text{is } \textbf{related to } \underline{\text{found phone}} \text{ AND} \\
&\quad \text{has } \textbf{value } "10" \text{ AND} \\
&\quad \text{has } \textbf{unit } "\%" \\
&\}
\end{align*}
\EndKnitrBlock{example}

\begin{figure}

{\centering \includegraphics[width=0.6\linewidth]{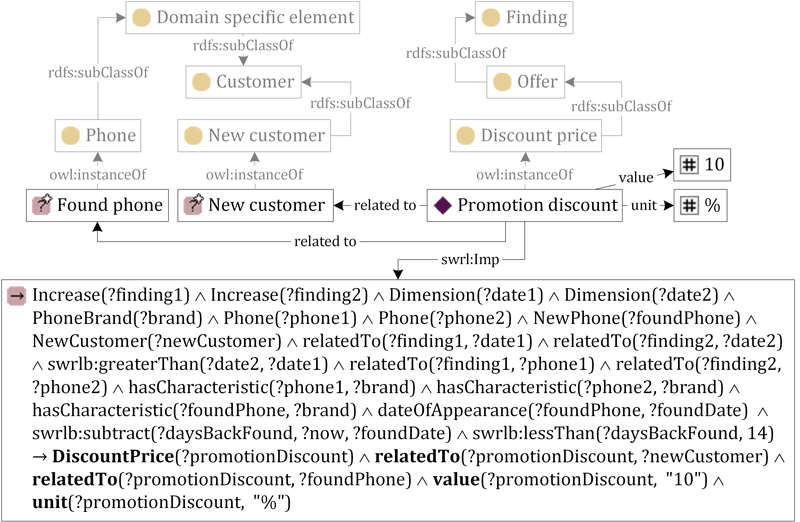}

}

\caption{Constraint presented in ontology in SWRL and OWL syntax}\label{fig:SWRL-OWL-example}
\end{figure}

When constraint presented in Example \ref{exm:template-rule} is
transformed to execution form at information system level, standardized
SWRL and OWL languages are used to enable reusability (see Figure
\ref{fig:SWRL-OWL-example}). By this transformation a rule is produced
that can be directly used in the inference engine to produce results in
a form of inferred triples that are presented to the user.

\subsection{Case study}\label{case-study}

\begin{figure}

{\centering \includegraphics[width=0.7\linewidth]{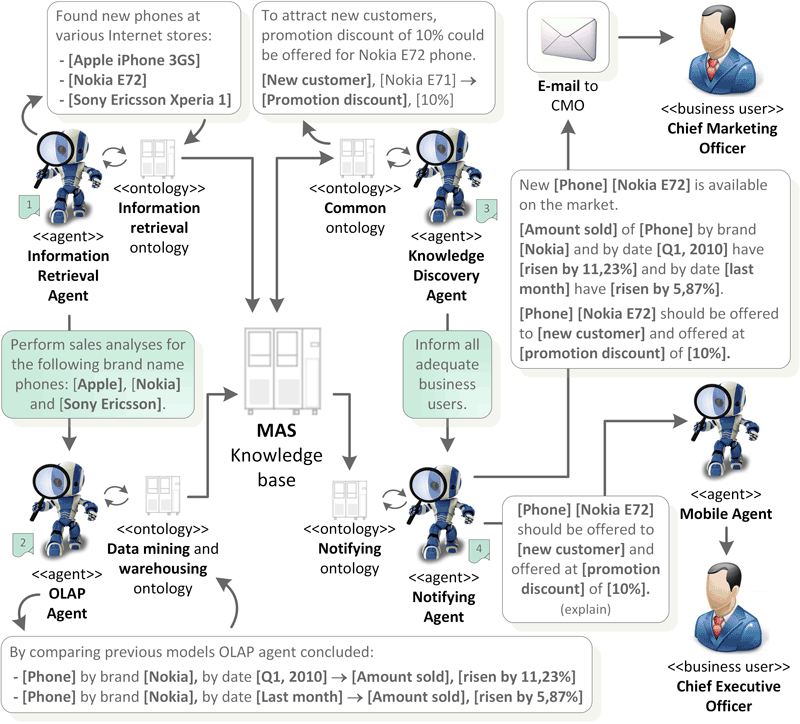}

}

\caption{Case study of using DSS-MAS in mobile phone domain}\label{fig:case-study}
\end{figure}

One of the most common cases of using DSS-MAS system is combing
information found online with BI reports (DW, DM, IR, etc.) developed on
internal data in enterprise DW. Figure \ref{fig:case-study} presents one
of this scenario.

Scenario presented at Figure \ref{fig:case-study} is triggered by
results of IRA activity, when three new mobile phones:
\texttt{Apple\ iPhone\ 3GS}, \texttt{Nokia\ E72} and
\texttt{Sony\ Ericsson\ Xperia\ 1} are found by IRA at online mobile
shops. According to the execution policy from Common ontology, OLAPA is
notified with a request to rebuild all DW reports where brands of
identified phones can be found in dimension elements. After running OLAP
on Sales schema with constrains of Nokia brand in Phone dimension and
last year in Date dimension OLAPA creates a report as depicted in
Example \ref{exm:bi-findings}.

\BeginKnitrBlock{example}[Business Intelligence findings]
\protect\hypertarget{exm:bi-findings}{}{\label{exm:bi-findings}
\iffalse (Business Intelligence findings) \fi{} }

\begin{align*}
&[ \textbf{Phone} ] \text{ by brand } [ \textbf{Nokia} ] \text{, by date } [ \textbf{Q1, 2010} ] \rightarrow [ \textbf{Amount sold} ] \text{, } [ \textbf{risen by } \boldsymbol{11,23\%} ] \\
&[ \textbf{Phone} ] \text{ by brand } [ \textbf{Nokia} ] \text{, by date } [ \textbf{Last month} ] \rightarrow [ \textbf{Amount sold} ] \text{, } [ \textbf{risen by } \boldsymbol{5,87\%} ]
\end{align*}
\EndKnitrBlock{example}

\begin{figure}

{\centering \includegraphics[width=0.6\linewidth]{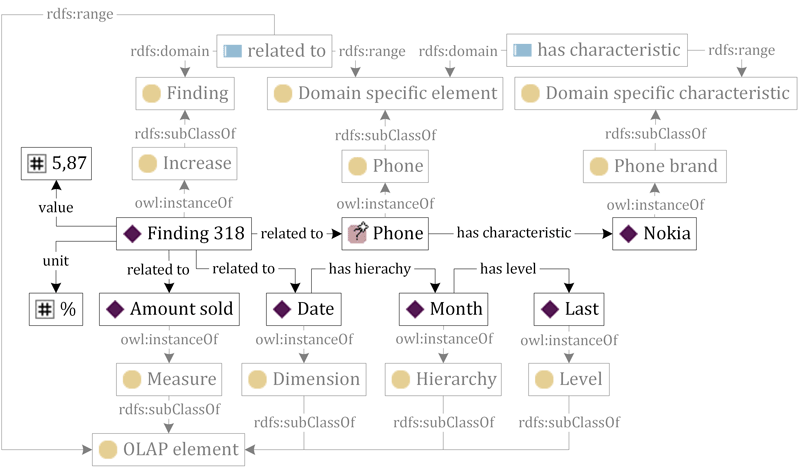}

}

\caption{Example of representing the finding in ontology}\label{fig:ontology-finding}
\end{figure}

At the information system level the first finding is represented as an
excerpt from ontology and is depicted in Figure
\ref{fig:ontology-finding}.

The fields that appear in the report are all instances of \textbf{Domain
specific element}, \textbf{OLAP element} and \textbf{Finding} from
ontology (see Figure \ref{fig:ontology-overview}). After these findings
have been asserted, KDA will be executed to derive new knowledge. Based
on these new facts represented at Example \ref{exm:bi-findings} and
enterprise business rules (see Example \ref{exm:template-rule}), the KDA
produced results represented in Example \ref{exm:derived-finding}, by
using inference engine knowledge is asserted in ontology.

\BeginKnitrBlock{example}[Derived finding]
\protect\hypertarget{exm:derived-finding}{}{\label{exm:derived-finding}
\iffalse (Derived finding) \fi{} }

\begin{align*}
&[ \textbf{New customer} ] \text{ , } [ \text{Nokia E72} ] \rightarrow [ \textbf{Promotion discount} ] \text{, } [ 10\% ]
\end{align*}
\EndKnitrBlock{example}

After consolidation of all new findings KDA sends message to NA with
request to forward notifications to appropriate users. The result of
triggered activity of NA is the list of business users that have to be
notified about this event. The list shows that in this case CMO and CEO
have to be notified whereas their context has to be considered.
According to CMO's preferences an e-mail is sent with the following
content presented in Example \ref{exm:findings-with-explanation}.

\BeginKnitrBlock{example}[Report of findings with explanation]
\protect\hypertarget{exm:findings-with-explanation}{}{\label{exm:findings-with-explanation}
\iffalse (Report of findings with explanation) \fi{} }

\begin{align*}
&\text{FACTS} \\
&\quad \text{New } [ \textbf{Phone} ] [ \textbf{Nokia E72} ] \text{ is available on the market.} \\
&\quad [ \textbf{Amount sold} ] \text{ of } [ \textbf{Phone} ] \text{ by brand } [ \textbf{Nokia} ] \text{ and by date } [ \textbf{Q1, 2010} ] \text{ have } [ \textbf{risen by } \boldsymbol{11,23\%} ] \\
&\quad \text{and by date } [ \textbf{last month} ] \text{ have } [ \textbf{risen by } \boldsymbol{5,87\%} ]. \\
&\text{CONCLUSION} \\
&\quad [ \textbf{Phone} ] [ \textbf{Nokia E72} ] \text{ should be offered to } [ \textbf{new customer} ] \text{ and offered at } [ \textbf{promotion discount} ] \\
&\quad \text{of } [ \boldsymbol{10\%} ].
\end{align*}
\EndKnitrBlock{example}

The CEO uses a Mobile Agent on his mobile device and is also notified by
a truncated message with new finding, while explanation is available
upon request.

\section{Conclusion}\label{conclusion}

In this paper we have discussed DSS-MAS where internal and external data
is integrated using agent-oriented approach and ontologies as a common
understanding of a problem domain and for communication between business
users and agents. Agents were used due to their mentalistic notions for
modelling, similarities between the agent in the MAS paradigm and the
human actor in business organisations, and also possibilities for the
use of ontologies as means of agents' internal knowledge base
representation. The external information from the Web was integrated by
IRA agent with the data in organisation's DW and after applying BR new
knowledge was derived by employing agents' inference capabilities. Tasks
like information retrieval from competitors, creating and reviewing OLAP
reports are autonomously performed by agents, while business users have
control over their execution through manipulation of knowledge base. The
research also has emphasized agent-to-business user communication,
trying to minimize that gap. This was accomplished by introducing
different views on ontologies for business user and agent. While agents
deal with formal description of business concepts, logical constraints
and rules, business user has a simplified view of formal description of
knowledge. User is able to manipulate with ontology through templates,
where little technical knowledge is required. The mediation mechanism
transforms these business level concepts into formal specification at
the level of information system.

Presented approach was verified and implemented using a case study from
the domain of mobile telecommunications, where the aim was to provide
the knowledge worker an intelligent analysis platform that enhances
decision making process. The application domain was reduced to its sub
domain dedicated for supply and analysis of demand of mobile phones in
one of the mobile operators. DW system is constructed from several
heterogeneous data sources where majority of those sources are internal
to the enterprise. Our approach added information found on the Web
(i.e.~competitors' offers, stock rates, etc.) to these internal data
sources and improved the decision support process within the enterprise.
The proposed approach also addressed business users and their
communication with the system which was simplified by using templates to
define some business requirements that were transformed into analyses
models (OLAP, DW, etc.), automatically performed by agents which
reported results back to users in charge. The case study presented in
the paper was implemented in Java and using mainly open source
technologies.

\end{document}